\documentclass[amsfonts, amssymb, amsmath, reprint,aip]{revtex4-1}
\usepackage[english]{babel}
\usepackage[utf8]{inputenc}
\usepackage[colorinlistoftodos, color=green!40, prependcaption]{todonotes}
\usepackage{mathtools}
\usepackage{physics}
\usepackage{xcolor}
\usepackage{graphicx}
\usepackage[left=23mm,right=13mm,top=35mm,columnsep=15pt]{geometry}
\usepackage{adjustbox}
\usepackage{placeins}
\usepackage[T1]{fontenc}
\usepackage{csquotes}
\usepackage{caption}
\usepackage{subcaption}
\usepackage{float}

\usepackage[pdftex, pdftitle={Article}, pdfauthor={Author}]{hyperref} 
\begin{document}
\title{Ground-state energy of quasi-free positrons in non-polar fluids}

\author{Eve Cheng}
\affiliation{Research School of Physics, Australian National University, Canberra}
\author{Daniel Cocks}
\affiliation{Research School of Physics, Australian National University, Canberra}
\author{Robert P. McEachran}
\affiliation{Research School of Physics, Australian National University, Canberra}

\newcommand{\Schrodinger}{Schr\"{o}dinger }
\newcommand{\col}{\color{black}}
\date{\today} 

\begin{abstract}
We have calculated the background energy ($V_0$) for positrons in noble gases with an \emph{ab initio} potential and the Wigner-Seitz (WS) ansatz. {\col In contrast to the general pseudo-potential approach, we have used accurate \emph{ab initio} potentials for the positron-atom interaction.} The ansatz includes an assumed form of the potential, resulting from an average over fluid atoms, and we propose four different options for this. By comparing the different options to literature data for effective electron number ($Z_{\text{eff}}$), we find that agreement can be obtained for light elements, but fails for heavy elements. We suspect that the strong polarizability of the heavy elements makes the simple potential averaging, as assumed in the Wigner-Seitz model, insufficient to fit the measurements without also making use of pseudo-potentials. We also raise our suspicion that the comparison of annihiliation rates between ground-state calculations and experimental values is not appropriate. Furthermore, the congruence of $V_0$ to $Z_{\text{eff}}$ values predicted by a contact potential approximation appears to be invalidated by our results.
\end{abstract}

\maketitle

Many technological applications involve the movements of electrons/positrons through a fluid, including medicine, industry and some environmental advances \cite{vanraes2018plasma,jones1982application,parker2008positron}. Hence, it is vital to understand the mechanisms behind the transport of quasi-free electrons/positrons, and how they can be accurately simulated \cite{cocks2019excitation}. While the positron/electron propagation has been extensively studied theoretically in dilute gases \cite{Iakubov1982,mceachran1979positron,mceachran1980positron,mceachran1977positron,mceachran1983elastic}, the theoretical basis for positron propagation in liquids is unsatisfactory. As most experimental comparisons have been made with electrons, positrons offer a unique opportunity to test the core assumptions and improve the predictive power of theoretical models. Therefore in this paper, we focus on the effects of density on the propagation of positrons in fluids.

The density dependence of the propagation of positrons in liquids is more challenging than in dilute gases. Because of the closer proximity in liquids, clusters or bubbles among other phenomena can occur because of the density-dependence of the charged particle/fluid interaction \cite{Rosenblit1997,Plenkiewicz1986,Nieminen1980,VanReeth1998,Iakubov1982}. This paper will focus on the propagation of the positron in homogeneous media, as understanding this intrinsic behaviour is necessary to understand those more complicated phenomena.

One noticeable density dependent property of a fluid is the non-negligible background contribution it makes to propagating states, often referred to as "quasi-free" states. The background contribution is evident in the lowest energy that a quasi-free particle can have, known as $V_0$ \cite{Iakubov1982}, sometimes referred to as the self energy \cite{green2014positron}. The values for $V_0$ can be on the order of a fraction of an eV below, or above, the vacuum energy. Because $V_0$ affects the resulting wavefunction of the particle, it is required for calculating its effective scattering behaviours, such as the annihilation rate for positrons. It also influences the ionisation potential of any atoms/molecules in fluids. In addition, $V_0$ provides the barrier at liquid-gas interfaces, which can lead to some interesting properties \cite{Borghesani2014, Evans2010}. Therefore, in this paper, we focus on the density dependence of $V_0$ for positrons in fluids of the noble-gas atoms ranging from gas to liquid densities.

For low density fluids, the linear density dependence of $V_0$ can be estimated by the optical pseudo-potential approximation \cite{Iakubov1982}. We will later show that this may be a bad approximation even in the limit of zero-density, where many-body effects could be expected to be negligible. In dense fluids, it is clear that the potential can no longer be simplified by a contact potential; the density dependence of $V_0$ is no longer linear.

There have been many experimental attempts to obtain $V_0$ for electrons in non-polar  fluids \cite{reininger1983relationship,Evans2010,Borghesani2014}. The most intuitive experimental approach is to measure the photoemission thresholds of a metal surface in vacuum and the target fluid \cite{reininger1983relationship,asaf1983energy}. However, this approach is limited because of the possible contamination and coating of the metal surface. Instead, Evans and co-workers \cite{Evans2010} proposed a different indirect approach that obtains $V_0$ by measuring the photoionization spectrum for the metal surface in liquids under two electric fields. They have measured $V_0$ for electrons in liquid Ar, Kr, He, Ne, $\text{N}_2$, $\text{CO}_2$, methane and ethane \cite{Shi2008,Evans2015,Lushtak2012,Shi2007,Lushtak2012,Evans2005,Evans2011,Shi2009}. Borghesani \cite{Borghesani2014} measured $V_0$ for electrons using the resonance attachment frequency of electrons to oxygen molecules. The $V_0$ values were calculated by comparing the attachment frequencies of the electrons in vacuum and target liquid. 

All experimental approaches for $V_0$ in fluids so far are limited to electrons only. There is no experimental data on $V_0$ values for positrons due to the difficulty of creating quasi-free positrons at low energies inside of a fluid. There was one study done on the $V_0$ calculations and measurements in solids \cite{plenkiewicz1996simple}, which was used to compare to the $V_0$ calculation in liquids in the study. Fortunately, there are some indirect approaches, which obtain $V_0$ by measuring the effective number of electrons the positron can annihilate with per atom in a fluid, commonly referred to as $Z_{\text{eff}}$ \cite{cocks2020positron}. 

There have been several existing theoretical approaches to calculate $V_0$ in fluids. The ``finite size" approach \cite{Chandler1994, Space1992} finds a solution locally within a finite system and then extrapolates to the macroscopic scale. The \Schrodinger equation was solved by mapping the electron wavefunction to an isomorphic polymer chain using Monte Carlo methods \cite{Chandler1994}. There are also other similar approaches that evaluate the electron wavefunction on a small grid and then extrapolate to larger systems \cite{Space1992}. These approaches are computationally limited, as only small systems can be simulated in this way{\col, although advances in numerical techniques have pushed this to an impressive scale\cite{cubero2003electronic}.}

The ``focus atom" approach was first used by Springett and coworkers \cite{Springett1968}. The approach uses the Wigner-Seitz (WS) model, which models the fluid as a collection of spherical unit cells centered around each liquid atom. In order to find $V_0$ by solving the \Schrodinger equation, they calculate the potential interaction between the electrons and the fluid by focusing on the interaction between the electron and a focus atom, and then calculating the total effective interaction as the ensemble average. {\col Since then a wide range of studies of the noble-gases and non-polar molecules have been performed, see Holroyd and Schmidt\cite{holroyd1989transport} and Iakubov and Khrapak\cite{Iakubov1982} and references therein for reviews of the original forays into this field. Many improvements on the pseudo-potentials used in the theory were made, along with techniques to include features of the interaction that could not be addressed by the pseudo-potentials\cite{boltjes1993computation, plenkiewicz1989density,plenkiewicz1988pseudopotential,plenkiewicz1988pseudopotential,Plenkiewicz1986}.
Recently, another advancement on this calculation method was published by Evans and co-workers \cite{Evans2010} later along with their experimental results.} This study is built upon the original WS model through introducing the concept of local WS cell (LWS) radius; it calculates $V_0$ for electrons as a combination of the ensemble polarization energy, the kinetic energy and the thermal energy. A series of calculations and experiments were performed \cite{Reininger1983, Evans2010} through a wide variety of species, including the noble gas atoms and simple hydrocarbons. The results showed significant agreement between the model and experimental data for electrons.

While the ``focus atom" approaches focus on finding reasonable approximations to the potential in the vicinity of one atom and taking an average of the rest of the atoms in the fluid, alternative techniques were also proposed. Some studies \cite{Polischuk1985,Lax1951} focus on the representation of the transition matrix, considering averages over multi-particle terms \cite{Lax1951} or diagrammatic approaches with vertex terms \cite{Polischuk1985}.

 This paper calculates $V_0$ using the ``focus atom" method and solves the \Schrodinger equation for the positron directly with both WS and LWS radius. The purpose of using both options for the radius is to test the extensibility of the LWS method by applying it to positrons, using accurate positron-atom potentials without free parameters. However, we also extend upon the previous approaches to include four different options for including the contribution to the potential felt by the positron from the atoms surrounding the focus atom. We contrast these approaches and their suitability.
 
 The structure of this paper is as follows: the first section is dedicated to the theory for our calculations. It goes through the \Schrodinger equation that we solve, the potential term in the equation and the corresponding boundary conditions used. The screening function and the calculation for $Z_{\text{eff}}$ are also discussed. The second section describes the numerical implementation using an in-house program PEEL (calculator of Positron and Electron Energy in Liquids). The third section goes through the results generated from the program and discusses the implications and potential future work and publications.


 %
 %
 %

\newcommand{\bb}[1]{\textbf{#1}}
\section{Theory}
The value of $V_0$ describes the lowest energy that a positron can have in a particular fluid. In this paper, we find $V_0$ by solving for an approximate solution to the \Schrodinger equation for the wavefunction of the positron in fluids with an ansatz. Our method is related to that of Springett {\it et al} \cite{Springett1968}, and Evans {\it et al} \cite{Evans2010}, although the derivation itself differs. 

In our approach, we first decouple the motion of the positron from the atoms by freezing the atomic positions, which is similar to Springett {\it et al}\cite{Springett1968}. Consequently, we aim to solve the single-particle \Schrodinger equation for the many-body system.

We solve this equation by taking an ansatz for the wavefunction. Same with the WS model, we define the liquid as a collection of WS cells. We then separate the liquid environment into two regions: inside or outside of the WS cells. Inside the WS cells, the potential and the corresponding ground-state wavefunction are spherically symmetric; outside the cells, the potential is $V_0$, and the corresponding wavefunction is constant:
\begin{align}\label{bc}
  \psi_{k_i}(\bb{r}) =
  \begin{cases}
    R(\bb{r}) & r < r_\mathrm{cell} \\
  \text{const} & r > r_\mathrm{cell}
  \end{cases}
\end{align}
The necessary boundary condition to connect the inside/outside regions is that the first derivative of the wavefunction at the edge of the WS cells is zero:
\begin{align} \label{BC}
    (\partial \psi_0/\partial r) |_{r = r_\mathrm{cell}} = 0
\end{align}
where the quantity $r_\mathrm{cell}$, the radius of the cell, can take two different values in this paper. The conventional choice is $r_\mathrm{cell} = r_{WS}$, where the Wigner-Seitz radius is given as
\begin{align} \label{WSradius}
    r_{WS} = (\frac{3}{4  \pi \rho_0})^{1/3}.
\end{align}
where $\rho_0$ is the atomic density of the fluids.

This effectively partitions the entire fluid volume into equal volumes for each atom. Based on the ansatz for the wavefunction, it is clear that the WS model assumes no overlap among WS cells. However, because the atoms in fluids fluctuate, this restriction of no overlap cannot be guaranteed physically. Therefore, it is our intention to acknowledge this limitation of the WS model and deal with the potential overlaps among WS cells; this is reflected in our definitions of the potential interaction later.

Another definition for the cell radius was proposed by Evans {\it et al} \cite{Evans2010}. Instead of partitioning the WS cells equally, the radius was adjusted to a smaller value, $r_\mathrm{cell} = r_{LWS}$, which takes account of the pair correlator in the liquid. The new radius is called the local WS (LWS) cell radius:
\begin{align} \label{RWS}
    r_{LWS} = (\frac{3}{4  \pi \rho_0 g_{max}})^{1/3}
\end{align}
where $g_{max}$ represents the maximum of the pair correlation function, $g(r)$ (discussed in further detail below).

For electrons, the LWS radius is shown to lead to reasonable agreement between theory and experiments in dense fluid systems \cite{Evans2010}, and does very well at predicting the behaviour around the critical point. We investigate both WS and LWS radius to observe the effects of different radii on the resulting $V_0$ values; this should also tell us whether the LWS radius can be extended to positrons or not.

Since the boundary condition requires the corresponding wavefunction to be spherically symmetric in each cell, we find $V_0$ by solving the single-particle \Schrodinger equation within a cell:
\begin{align} \label{SE}
  [\frac{d^2}{dr^2} - U(r) +k^2]u(r) = 0
\end{align}
where $\frac{\hbar^2}{2m}U(r)$ is the total positron-atom interaction, $k$ is the wavenumber and $u(r)/r$ is the total radial wavefunction.

We note here that while the WS model is sufficient in calculating the ground-state eigenvalue, it does not provide the infrastructure to calculate higher energy states with non-zero momentum. The boundary condition of WS model forces the wavefunction to be spherically symmetric, which does not accommodate higher energy states. Our future goal is to calculate the bandstructure of the liquid; it would require a completely new approach, and it will be discussed in future work.

%
%
%

\subsection{The total potential in fluids}
The difference between our approach and the traditional WS approach \cite{Springett1968} lies in our potential term. Our total potential for the interaction between the positron and all atoms is treated as precisely as possible for one atom (referred to as the ``focus atom"), and averaged for all others. In terms of the interaction between the focus atom and the positron, instead of using pseudopotentials as most of the WS approaches \cite{Evans2010,plenkiewicz1996simple,Springett1968}, we use an \emph{ab initio} potential, which has been verified against single scattering cross sections \cite{boyle2015electron,McEachran1977}. There are also modifications implemented to the averaging of the potential, which will be elaborated in this section.

In short, our total potential can be written as:
\begin{align}\label{totalU}
    U = U_1 + U_2
\end{align}
where $U$ is the total potential, $U_1$ is the contribution of the focus atom, including screening effects, and $U_2$ is the contribution from the surrounding atoms.

We define the focus atom as an atom within a neighbourhood of the WS radius of the positron; this neighbourhood changes as the positron moves through the fluid. Because we acknowledge the overlap among WS cells, there might not be a unique choice for the focus atom; this ambiguity will be addressed later when defining $U_2$.

\subsubsection{Focus atom interaction with the positron $U_1$}
\begin{figure*}[t]
     \includegraphics[width=0.5\textwidth]{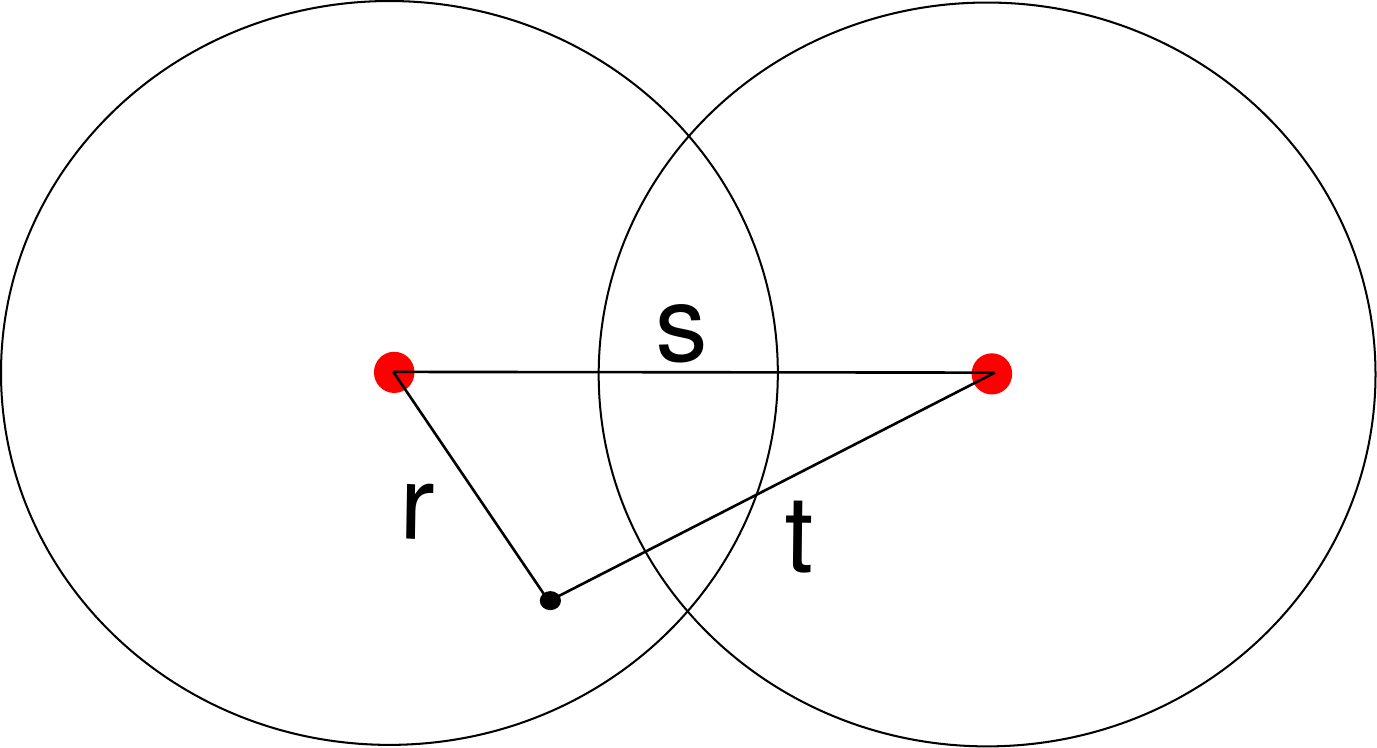}
     \caption{This diagram depicts our method for calculating the average potential interaction between the positron and the liquid atoms. The red dots represent the atoms, while the black small dot represents the positron. The left red dot is the focus atom, while the right red dot is a randomly selected neighbouring atom. The variabless $r$, $s$, $t$ are consistent with equation (\ref{fr}) and the definitions for the interaction potentials.). \label{diagram}}
\end{figure*}
The interaction between a positron and an isolated atom ($V$) can be described using the polarized orbital method \cite{McEachran1977}, which separates the potential into two parts: the static ($U_s$) and the polarization potential ($U_p$). The static potential represents the Coulomb interactions with the unperturbed Dirac-Fock orbitals, whereas the polarization potential is given by a multipole expansion up to octopole terms in the perturbation of the orbitals. Due to the lack of exchange interaction, the representation of the polarisation interaction in positron scattering is more important than in electron scattering.

In a liquid, the interaction between the positron and the focus atom ($U_1$) can only be partially described by ($V$) above. In addition to $V$, we must account for screening \cite{Lekner1967}. The basic element of screening involves the cooperation between each pair of atoms, as depicted in Figure \ref{diagram}. The variables $r$, $s$, and $t$ refer to the positron-focus atom, atom-atom, and positron-neighbouring atom distances respectively. With all considered, $U_1$ can be expressed as:
\begin{align}
  U_1 = U_s + f(r) U_p,
\end{align}
where $f(r)$ is a screening function.

Assuming that the largest contribution to screening comes from the dipole-dipole interaction between the atoms, we may use the well-known \cite{Lekner1967} self-consistent expression for $f(r)$:
\begin{align}\label{fr}
  f(r) = 1 - \pi n \int^{\infty}_0 ds ~ s^{-2} g(s) & \int^{r+s}_{|r-s|}dt t^{-2} f(t)\\ &\alpha(t) \Theta (r,s,t) \nonumber
\end{align}
where $g(s)$ is the radial distribution function, and $\alpha(r)$ is the dipole polarisability as a function of $r$, which is the dominant contribution to $U_p$ and,
\begin{equation}
  \Theta = \frac{3}{2} s^{-2} (s^2+t^2-r^2)(s^2+r^2-t^2) + (r^2+t^2-s^2).
\end{equation}
We obtain $f(r)$ by iterating \eqref{fr}, starting from a constant value $f(r)=f_L$ where $f_L$ is the Lorentz local field:
\begin{align}
    f_L =1/(1 + (8/3)\pi n \alpha),
\end{align}
and $\alpha=\alpha(r\rightarrow\infty)$ is the asymptotic dipole polarisability.

{\col While the most obvious effect that the surrounding atoms have on the focus-atom-positron interaction is the polarization reduction from static screening, there are other effects that we have neglected. For example, dynamic polarization is difficult to handle even for the case of single-atom positron scattering, so we do not include the effects of surrounding atoms on the focus-atom dynamic polarizability.}

\subsubsection{The effective potential between the surrounding atoms and the positron $U_2$}
As mentioned above, the acknowledgment of the overlap among WS cells has certain implications when defining $U_2$. Because of the existence of overlaps, there can be more than one atom within the neighbourhood of WS radius centering the positron. If no other limitations are applied, the interaction between a positron and the liquid environment (denoted as $U_2^T$) can be calculated as an ensemble average of the single atom interaction potential:
\begin{align}\label{total_}
  U_2^T(r) &= \frac{2\pi \rho}{r} \int_0^{\infty}dt\ t U_1(t)\int^{r+t}_{|r-t|}ds\ s g(s),
\end{align}

The $U_2^T$ option for $U_2$ assumes no correlation between the positron and the surrounding atoms and relies solely on the atom-atom correlation $g(r)$. This is the simplest choice because it averages over the positron-surrounding atom separation. It effectively assumes a Born-approximation-like interaction between the positron and surrounding atom. Hence, it includes large unphysical contributions from the divergent static potential.

While still enforcing no additional constraints on the identity of the focus atom, the second option ($U_2^P$) for $U_2$ takes only the ensemble average of the polarization potential $U_p$ to avoid avoid the divergent spike of the static potential as $t\rightarrow 0$:
\begin{align}
  U_2^P(r) &= \frac{2\pi \rho}{r} \int_0^{\infty} dt\ t U_p(t)\int^{r+t}_{|r-t|} ds\ s g(s) 
\end{align}
The $U_2^P$ option assumes the atoms are far away from each other enough that the static potential coming from non-focus atoms does not contribute to the interaction, although it does allow short-range contributions from the polarisation potential. This tends to over-emphasise the potential interactions and therefore serves as a lower bound to $V_0$. 

Our third and fourth options ($U_2^r$ and $U_2^{U_1}$) apply a cut-off radius, $r_0$ when calculating $U_2$. The general form of $U_2$ for these two options is:
\begin{align}\label{cutoffwithr}
  U_2^R(r;r_0) &= \frac{2\pi \rho}{r} \int_{r_0}^{\infty} dt\ t U_1(t)\int^{r+t}_{|r-t|} ds\ s g(s)
\end{align}
The application of a cut-off radius applies additional constraint on the identity of the focus atom. The only difference between equation (\ref{total_}) and (\ref{cutoffwithr}) is that equation (\ref{cutoffwithr}) integrates from $r_0$ instead of 0 in the outer integral (i.e. $U_2^T(r) = U_2^R(r;0)$). This seemingly minor difference marks the focus atom to be the closest atom to the positron, acting as a second barrier upon $g(r)$ to keep atoms at reasonable distances from the positron. This is a reasonable choice for dense fluids, where the close packing forces this behaviour.

The third option ($U_2^r$) takes the choice of $r_0=r$:
\begin{align}
    U_2^r(r) = U_2^R(r;r).
\end{align}
In comparison, the fourth method ($U_2^{U_1}$) sets $r_0=r_{U_1}$, where $r_{U_1}$ is the radius where $U_1$ first crosses the r axis, i.e. $U_1=0$:
\begin{align}
    U_2^{U_1}(r) = U_2^R(r ; r_{U_1}).
\end{align}
This method assumes the positron is kept away from the core of the surrounding atoms by the repulsive static part of the potential. This method captures the main effects of $U_2^P$ with some additional repulsive interactions included.

The implementation of the LWS radius method for the potential is the same as the WS radius method, except $V_{WS}$ is defined as the spherical space with radius of the local WS radius instead.

A cut-off radius was often implemented in many existing WS approaches \cite{Evans2010,plenkiewicz1996simple,Springett1968} in a very different manner than ours. In the existing approaches, they use the sum of the hard-sphere static potential and the polarisation potential with a pre-defined cutoff radius to calculate $U_1$ \cite{Evans2010,plenkiewicz1996simple,Springett1968}. Different from those approaches which apply cut-off radii to the total potential, we use the full $U_1$ and only apply cut-off radii to $U_2$. We believe cut-offs in this fashion can potentially lead to more accurate results.

\subsection{The effective charge $Z_{\text{eff}}$}
Because there is no experimental data for $V_0$ values in the liquid phase, instead, we can draw a comparison between the effective charge, $Z_\text{eff}$, of our calculated ground state, and measured annihilation rates. The effective charge is calculated through determining the wavefunction overlap between the charged particle and the electron cloud around the focus atom in the liquid \cite{cocks2020positron}.
\begin{align}\label{zeff1}
  Z_{\text{eff}} = \sum_{i=1}^{N} &\int |\Psi(\mathbf{r_1},\mathbf{r_2},...,\mathbf{r_{N}};\mathbf{x})|^2 \delta(\mathbf{r_i}-\mathbf{x}) \\
  &d\mathbf{r_1}d\mathbf{r_2}...d\mathbf{r_{N}} \nonumber
\end{align}
where $\Psi$ is the total wavefunction of electron with $\mathbf{x}$ being the position vector of the positron, and $\mathbf{r_i}$ refers to the coordinates of the atomic electrons (spin included). 

The $Z_{\text{eff}}$ can also be written as \cite{cocks2020positron,mceachran1977positron}:
\begin{align}\label{zeff}
    Z_{\text{eff}} = Z_{\text{eff}}^0 + Z_{\text{eff}}^1
\end{align}
where 
\begin{align}
    Z_{\text{eff}}^i =  \int_0^{\infty} d\mathbf{r} |\psi(\mathbf{r})|^2\rho_i(\mathbf{r})
\end{align}
where $\rho_0(\mathbf{r})$ represents the unperturbed charge density of the atomic orbitals, and $\rho_1(\mathbf{r})$ refers to the first-order correction due to the polarization interaction between the electrons and the positron. It can be easily shown that equation (\ref{zeff}) is equivalent to equation (\ref{zeff1}).

We also extend the definition of the $Z_{\text{eff}}$ to include the electron density contribution from the surrounding atoms \cite{cocks2020positron}:
\begin{align}
    Z_{\text{eff}} &= \frac{1}{n} \int_0^{r_m} d \mathbf{r} (\rho_L(\mathbf{r}) + \rho_S(\mathbf{r}))|\Psi(\mathbf{r})|^2 \\
    &= Z_{\text{eff}}^L + Z_{\text{eff}}^S
\end{align}
Note that $Z_{\text{eff}}^L$ here is equivalent to the $Z_{\text{eff}}$ in equation (\ref{zeff}). Our calculation shows the contribution from $Z_{\text{eff}}^S$ is negligible for all gases.

\subsection{The averaged effective charge $\langle Z_\text{eff} \rangle$}
All of our calculations so far assume $k$ (the wavenumber) $\rightarrow 0$, however, this is not realistic for experimental measurements. In order to make some better comparisons with experiment, we also performed preliminary calculations of the thermally averaged $Z_{\text{eff}}$, i.e. $\langle Z_\text{eff} \rangle$.

We assume the thermal distribution for the positron is a Boltzmann distribution, and therefore the thermal average for $Z_{\text{eff}}$ can be represented by:
\begin{align}
    \langle Z_\text{eff} \rangle &= \int_0^{\infty} dE ~ W_{b}(E) \times Z_{\text{eff}}(E)\\
    W_b(E)&= N \times \sqrt{E} \exp{-E/kT}\\
    N &= \int_{0}^\infty dE ~ \sqrt{E} \exp{-E/kT}
\end{align}
where $W_b(E)$ is the Boltzmann weight factor for each thermal energy, $Z_{\text{eff}}(E)$ is the $Z_{\text{eff}}$ value for the thermal energy $E$, and $N$ is the normalization factor for the Boltzmann weight factor. In our numerical implementation, we truncate the integration over $E$ at an upper limit of $15kT/2$ rather than infinity.

\section{Numerical Implementation (PEEL)}
We have developed an in-house program called PEEL: calculator of Positrons and Electrons Energy in Liquids.
\subsection{Inputs for the Potential Calculation}

For atomic species, a tabulated set of values for $U_p$ and $U_s$ has been calculated, using the Dirac-Fock orbitals of the atom, as described in \cite{mceachran1977positron} but with updated parameters. We have calculated the screening function, $f(r)$, iteratively from equation (\ref{fr}) using the radial distribution functions, and the dipole polarisibility $\alpha (r)$ of the potential calculation.
\subsubsection{The radial distribution function $g(r)$}
The radial distribution function $g(r)$ is calculated iteratively solving the Ornstein-Zernike relation with Percus-Yevick approximation \cite{hansen1990theory}:
\begin{align}
  \begin{split}
  \exp[\beta v(r)] g(r) = &1 + \rho \int [g(r-r')-1]g(r')\\
  &(1 - \exp[\beta v(r')])dr'
\end{split}
\end{align}
where $v(r)$ is the pair potential between the liquid atoms and $\beta$ represents the thermal energy. 

We simplified the pair potential between two liquid atoms to be the Lennard-Jones potential for the $v(r)$ calculation:
\begin{align} \label{LJ}
  U(r) = 4 \epsilon \left[ \left(\frac{\sigma}{r}\right)^{12} -  \left(\frac{\sigma}{r}\right)^{6} \right]
\end{align}
where $U(r)$ represents the interatomic potential energy, $\epsilon$ is the depth of the potential well (the liquid atoms), and $\sigma$ is the distance at which the interatomic potential interaction becomes zero. The values for $\epsilon$ and $\sigma$ were generated through fitting for viscosity data of different fluids \cite{oh2013modified}. This is a commonly used model potential, which approximates the hard-core repulsion and van-der-Waals interaction of real atomic interactions.

An empirically deduced form of the reduced temperature was used in solving the Ornstein-Zernike relation \cite{oh2013modified}.
\begin{align} \label{reducedtemp}
 T^* = \frac{k_B (T - \tau)}{\epsilon}
\end{align}
where $k_B$ is thermal energy, and $\tau$ is the correction factor that is also generated through fitting the viscosity data. 

We have also checked the results using a molecular dynamic package Large-scale Atomic/Molecular Massively Parallel Simulator (LAMMPS) at selected densities and temperatures \cite{plimpton1995fast}. The iterative function is preferred because it is significantly less computationally expensive compared to LAMMPS.

\subsection{Scattering equation solver}
With the potential implemented, the program solves the \Schrodinger equation for the ansatz given in equation (\ref{BC}) with the lowest energy eigenvalue that satisfies the boundary condition.

With the potential term implemented using equation (\ref{totalU}), PEEL was set up to solve for the wavefunction with a given energy from $r$ = 0 to $r$ = $r_m$ (where $r_m$ is the cell radius). This uses the Julia package DifferentialEquations with callbacks to handle divergent behaviour in the classically forbidden region. Once the wavefunction is calculated, the program will provide both the values of the wavefunction and the derivative of the wavefunction.

The most intuitive way to ensure the boundary condition for equation (\ref{SE}) would be to run the wavefunction solver for a range of energies and record the one that satisfies the WS boundary condition with the expected number of turning points. The turning points refer to the points on the wavefunction where its derivative is zero, and $V_0$ corresponds to the wavefunction with only one turning point. This method can be used but tends to be very slow and relies heavily on the grid of input energies. 

However, to speed up the process and improve the accuracy of calculating $V_0$, a root-finding method was implemented instead of the intuitive one. The root-finding method searches for the root of a turning-point-counting function, which counts the number of turning points in a wavefunction within a range of energies. In order to increase the accuracy and speed, the range of energies is narrowed down first. The upper-limit is set as the lowest energy eigenstate for bound-state wavefunction. The lower-limit was calculated by finding out an energy corresponding to the wavefunction with no turning points. Once the energy range is calculated, we search for the root of the turning-point-counting function within the energy range calculated. The resulting root is $V_0$.

%
%
%
%

\section{Results}
The result section is structured as follows: we first focus on the four options for the form of $U_2$ and use argon as a test case to explore the density dependency of both $V_0$ and $Z_{\text{eff}}$ for each option. We also discuss the effects of the inclusion of the Boltzmann thermal distribution of the incoming positron to the calculation for argon. We then select one of these methods to explore the other noble gases: helium, neon, krypton and xenon. 

\subsection{Density dependence of $V_0$ for argon} 
\begin{figure*}[t]
  \begin{subfigure}[h]{0.45\textwidth}
     \includegraphics[width=\textwidth]{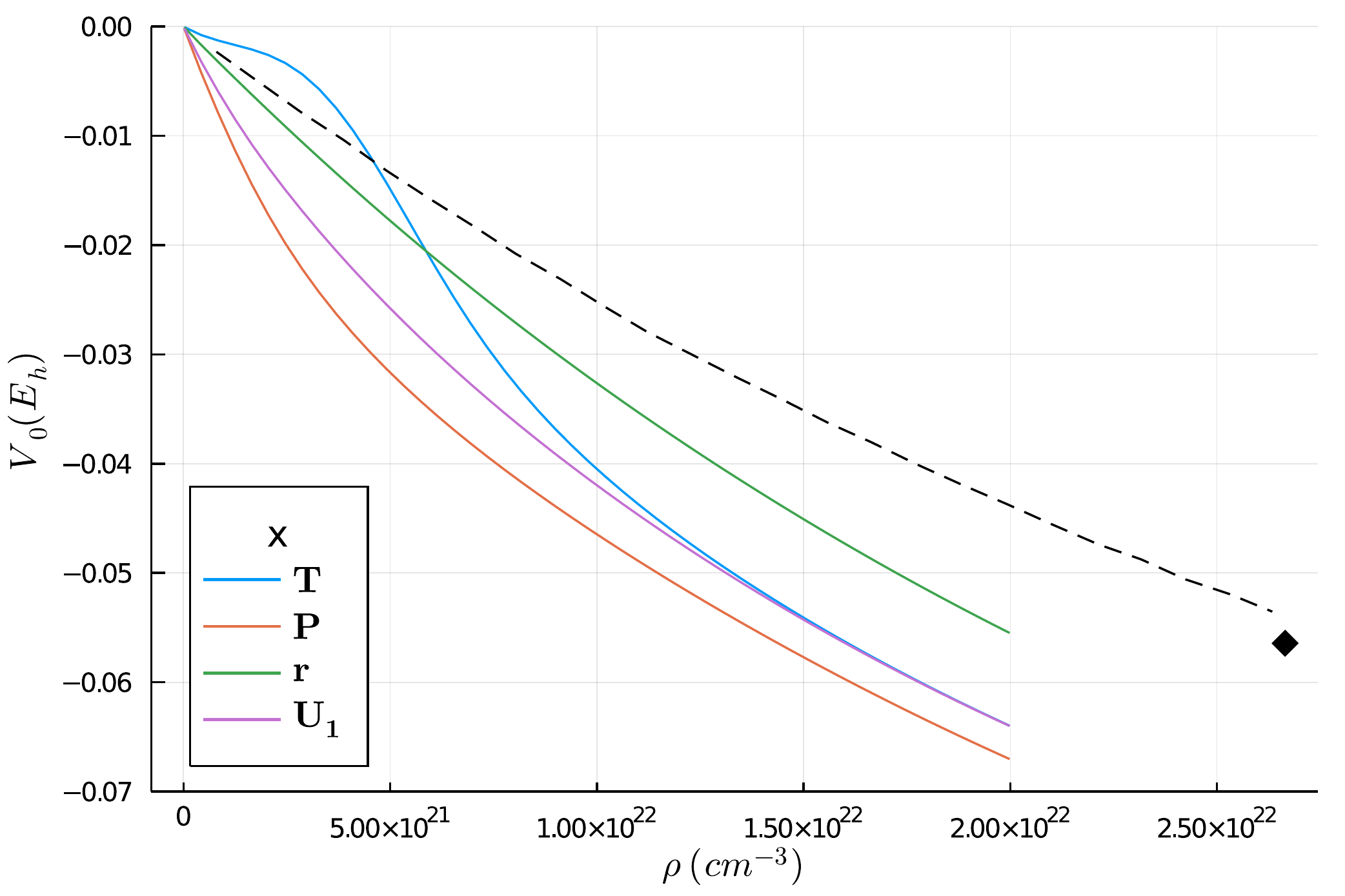}
     \caption{}
  \end{subfigure}
    \hspace{30pt}
  \begin{subfigure}[h]{0.45\textwidth}
     \includegraphics[width=\textwidth]{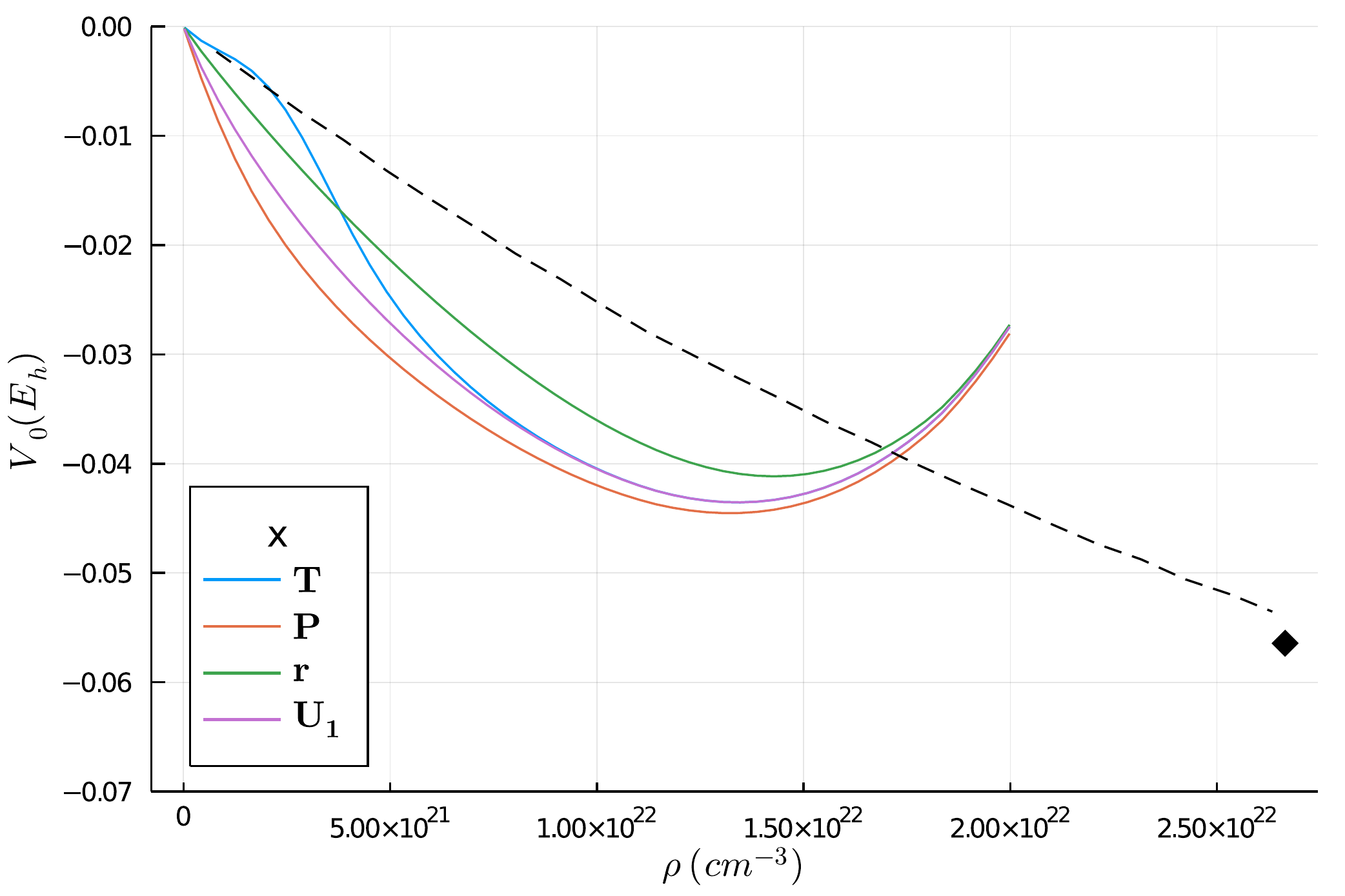}
     \caption{}
  \end{subfigure}
    \hspace{30pt}
\caption{\label{fig:argon-V0-Zeff}The density dependence of $V_0$ for argon at 295K. (a) $V_0$ calculation with $r_\mathrm{cell} = r_\mathrm{WS}$ (b) $V_0$ calculation with $r_\mathrm{cell} = r_\mathrm{LWS}$. The lines correspond to the different options for $U_2^x$, where x is shown in the legend. The black dashed lines in (a) and (b) come from the theoretical studies \cite{plenkiewicz1996simple} using a simple pseudo-potential comprised of hard-sphere core and dipole polarization terms. The black diamonds in (a) and (b) refer to the experimental results obtained in crystalline samples at 4 K \cite{gullikson1988observation}. }
\end{figure*}

Our results for the $V_0$ values of positrons in fluid argon using the different averaging options are shown in figure \ref{fig:argon-V0-Zeff}a) for the choice of $r_\mathrm{cell} = r_\mathrm{WS}$. A similar comparison is made in figure \ref{fig:argon-V0-Zeff}b) but for $r_\mathrm{cell} = r_\mathrm{LWS}$.

In general, the qualitative behaviour is very similar for the different options of averaging, i.e. the contribution to the potential from $U_2$. All $V_0$ values are negative, which is consistent with the scattering length of positron scattering from argon (see table \ref{table}). As the density increases, $V_0$ steadily increases in magnitude, which means the interaction between the positron and the fluid becomes more attractive with higher densities; this is also expected.

In the limit of low densities, a linear dependence on density for $V_{0}(\rho\rightarrow0)$ is observed in our results (see Figure \ref{fig:argon-V0-Zeff}). {\col This linearity is consistent with Iakubov and Khrapack's derivation \cite{Iakubov1982}, which uses a contact potential approximation for the positron-atom interaction:}

\begin{align}
    V(r)\approx\frac{2\pi\hbar^{2}}{m}a_{s}\delta(r)
\end{align}

Under the contact potential approximation, $V_0$ depends linearly on the density and the scattering length, $a_s$:
\begin{align} \label{scatterlength}
    V_{0}\approx\frac{2\pi\hbar^{2}}{m}a_{s}\rho
\end{align}

Since the scattering lengths are usually readily available, the slope of $V_0 (\rho\rightarrow0)$ is commonly used as a benchmark for low density regions (see Table \ref{table}).


 Compared to the contact potential approximation, our potential ($U(r)$) for the positron-atom interaction is calculated in a more precise manner. While $U_1$ should be consistent with the scattering length, one can view the four options provided for $U_2$ as alternatives to the contact potential approximation. {\col
 In this way, we can characterise the combined effects of the dilute fluid on the positron via a single quantity: $\lim_{\rho\rightarrow0} dV_0/d\rho$. One can draw an analogy between this quantity and the scattering length, which characterises single-atom scattering for low energy, as both are directly relevant to experiments in a particular limit and both result from the full set of interactions in the system. Similarly, the contact potential is the ``idealized case" for both of these quantities, and in a fluid this can be seen in the linear dependence in equation \eqref{scatterlength}. It should be pointed out, however, that our use of a ranged potential means $dV_0/d\rho$ will quickly deviate from its value in the dilute limit as denser fluids are considered.
 }
 
 At low densities, the inclusion of $U_2$ causes the linear trend of $V_{0}(\rho\rightarrow0)$ to deviate from the contact potential approximation; it preserves the linearity of $V_0 (\rho \rightarrow 0)$ but changes its gradient. The two choices $U^T_2$ and $U^P_2$ under- and over-estimate this low-density trend, and therefore serve as clear upper/lower bounds for $V_{0}(\rho\rightarrow0)$. As the upper bound,  $U^T_2$ weights the entire repulsive static potential equally, even for $r\rightarrow0$ where the positron wavefunction is suppressed, and consequently underestimates the slope. As the lower bound, $U^P_2$ completely neglects the static potential and therefore overestimates the slope. Inbetween two bounds, the option $U_2^r$ potentially underestimates the contribution from the polarisation potential interaction and consequently could result in  overestimation of the slope. The option $U_2^{U_1}$ produces $V_0$ values between $U_2^P$ and $U_2^r$ at low densities.

As the density increases in the low density region, the positron becomes more confined in the fluid, which in turn leads to a strong kinetic energy contribution. Because of the increasing contribution from the repulsive interaction between the positron and the nuclei at higher densities, the resulting $V_0$ values are expected to decrease with a slower rate. This expectation is met with all methods except $U_2^T$, because it over-estimates the repulsive interaction at lower densities. 

We expect that at sufficiently high densities, the repulsive interaction dominates the positron-fluid interaction and leads to an increasing trend in $V_0$. This appears in the case of $r_\mathrm{cell}=r_{WS}$ as a flattening of the slope, and is much more dramatic for the $r_\mathrm{cell} = r_{LWS}$ case due to the smaller effective sphere allowed for the positron. In fact, $r_{LWS}$ provides a turning point in $V_{0}(\rho)$. The existence of a turning point allows the potential to describe fluctuation clustering in fluids, where a positron can cause the a local fluctuation in density towards the point of $dV_{0}/d\rho=0$. This can be seen, for example, with clustering in helium caused by positrons; it is observed that clusters of fixed density form independently of the background fluid density \cite{Hautojarvi1977}. 

While there is a difference in the density dependence of $V_0$ at high densities with respect to the choice of $r_{\text{cell}}$, the relative position of $V_0$ curves using four different methods is roughly the same. The $U_2^P$ method still acts as the lower bound for $V_0$ calculations. Interestingly, $U_2^r$, instead of $U_2^T$, serves as the upper bound at high densities, for both $r_{\text{cell}}$ choices. This switch is due to the different sources for the overestimation of $V_0$ for these two methods. $U_2^T$ overestimates $V_0$ through overestimation of repulsive interaction, while $U_2^r$ underestimates the polarisation interaction. In addition, we have observed that the calculations from $U_2^{U_1}$ tend to converge to $U_2^T$ at high densities; we think this is because $r_{\text{cell}}$ is small enough at high densities that the positron is not allowed to be too close to other atoms even with option $U_2^T$. 

Among the four averaging choices, the $U_2^T$ choice shows an unusual feature by having initially a smaller slope at low densities. As previously discussed, this is likely due to its indiscriminate inclusion of the entire static potential and suggests that it is a poor choice for low densities. Note that we have used this choice in the past for cross section calculations of electron scattering in fluids \cite{boyle2015electron, cocks2020positron} as it is the simplest description of the surrounding average.

The results from previous theoretical calculations by Plenkiewicz \emph{et al} \cite{plenkiewicz1996simple} are shown in Figure \ref{fig:argon-V0-Zeff} as comparison for our calculations. Overall, their $V_0$ results are consistently higher than our results. At low densities, the slopes of their $V_0$ calculations match the scattering length reasonably well. However, we suspect that while the scattering length provides rough guidelines for $V_0$ at low densities, it is not a reliable metric for judging the accuracy of the $V_0$ calculation (discussed further in section \ref{sectionallgas}).  Moreover, their calculation requires the input of the scattering length and makes more assumptions regarding the potential interaction. Therefore, we believe the differences come from both the representation of the pseudo-potential, which has an effect on both the positron-atom interaction and the form of the ansatz (equation (\ref{bc})). 

\begin{figure*}[t]
 \begin{subfigure}[h]{0.45\textwidth}
     \includegraphics[width=\textwidth]{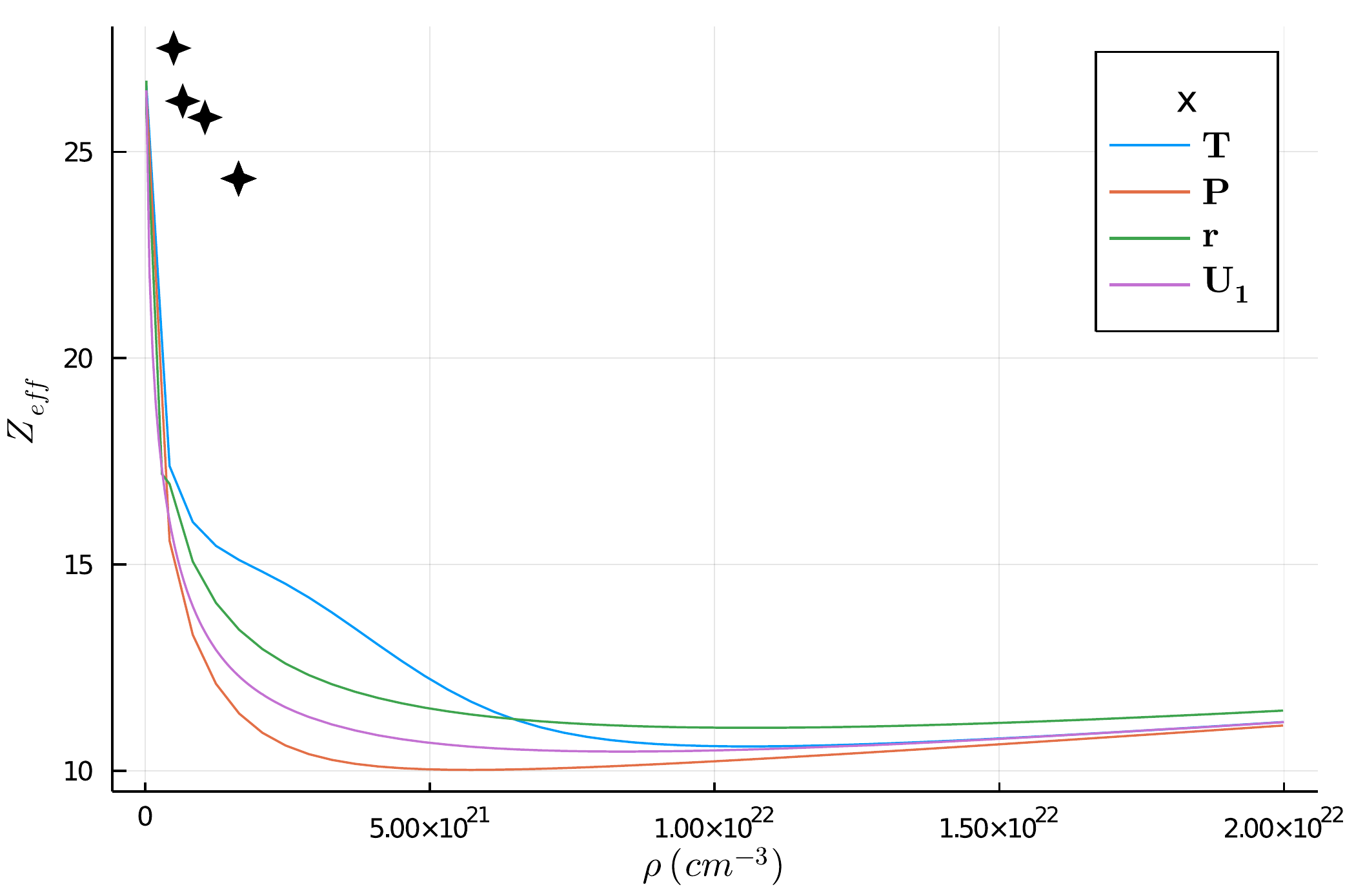}
     \caption{}
  \end{subfigure} 
   \begin{subfigure}[h]{0.45\textwidth}
      \includegraphics[width=\textwidth]{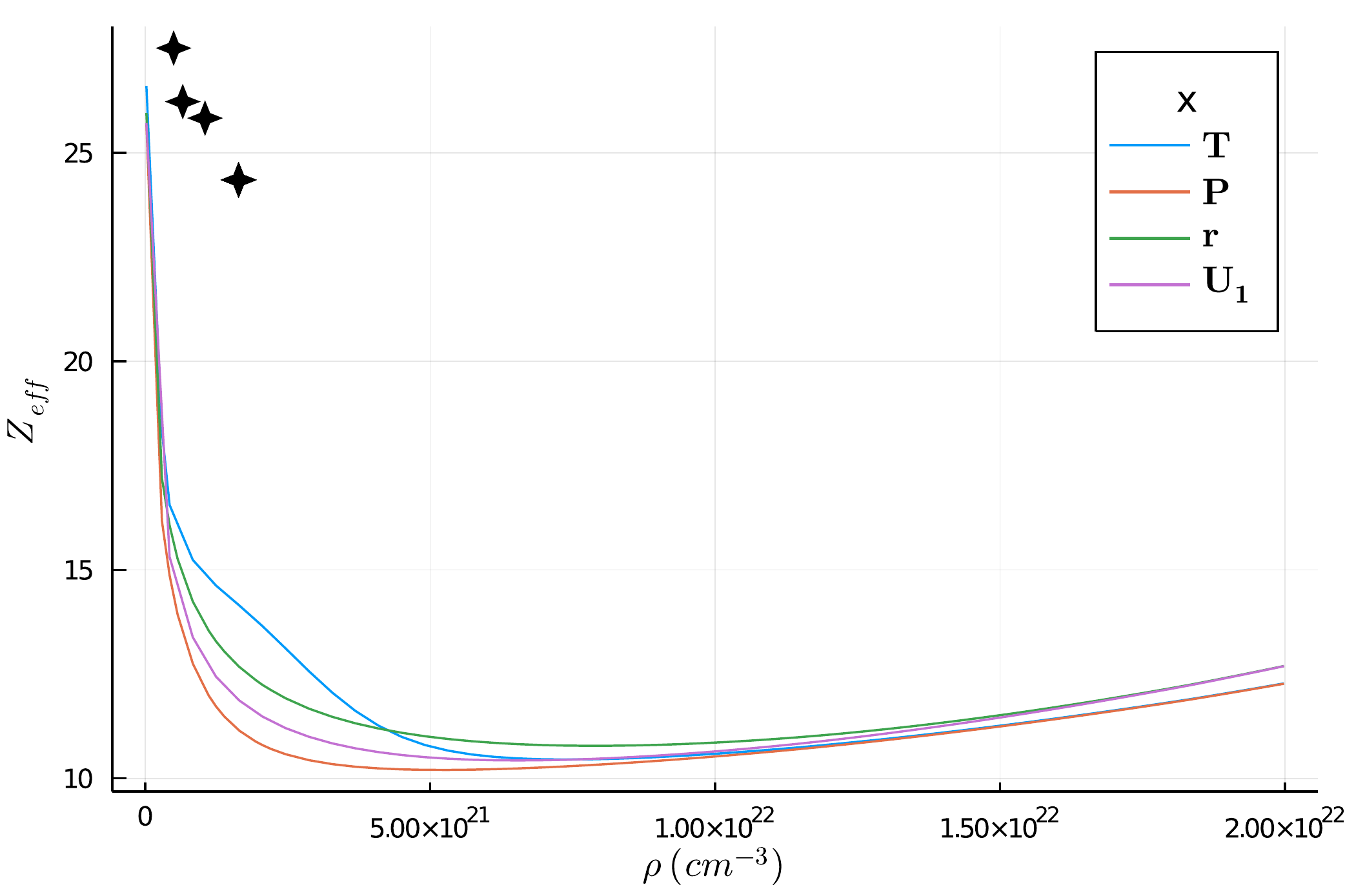}
     \caption{}
  \end{subfigure} 
 \caption{\label{fig:zeff-argon}(a) $Z_{\text{eff}}$ calculation with $r_\mathrm{cell}=r_{WS}$  (b) $Z_{\text{eff}}$ calculation with $r_\mathrm{cell}=r_{LWS}$ . The different lines correspond to the different options for $U_2^x$, where x is shown in the legend. The black stars in (a) and (b) refer to the experimental data for argon at 295K.\cite{tuomisaari1985localized}}
\end{figure*}

\subsection{Density dependence of $Z_\mathrm{eff}$ for argon}

It is difficult to measure $V_{0}$ for positrons in a fluid directly; the only related measurements we are aware of are for positrons in the solid rare gases \cite{plenkiewicz1996simple}. Fortunately, there are many measurements at different densities for the annihilation rate of the rare gases, which is proportional to $Z_{\mathrm{eff}}$. Hence, we have calculated $Z_{\mathrm{eff}}$ values for the lowest-energy positron states by using the wavefunction density and its overlap with the atomic orbital densities, and we have compared the results with experimental measurements.

Our calculations for the $Z_\mathrm{eff}$ values of the positron ground state are shown in figures \ref{fig:zeff-argon}a) and b) for $r_\mathrm{cell}=r_{WS}$ and $r_{LWS}$ respectively. The comparison with literature data shows disagreements with our calculation in the density dependence of $Z_{\text{eff}}$. While our calculation matches with literature data at low densities, it seems that the prediction for the ground-state is more sensitive to the density than the experimental measurements of the ensemble average.

The disagreement potentially comes from the fact that this comparison between our calculated $Z_{\text{eff}}$ values and the experimental data is not direct. The experimental measurements are an ensemble average of a distribution of positron kinetic energies, while the WS approach only calculates for the lowest energy state. Ideally, the experimental measurements result in a thermal distribution, however, it is often not the case, especially with heavier rare gases \cite{green2017positron}. The heavier rare gases have larger $Z_{\text{eff}}$ and slower thermalisation rates, therefore they do not allow the incoming positron beam to thermalise \cite{tuomisaari1988positron,wright1985annihilation}. We expect that the theoretical prediction for the energy shifts of the higher-kinetic-energy states to be smaller in magnitude than $V_0$, because they are able to move more freely throughout the fluid. Therefore, the difference between experiment and theory might be exaggerated in this comparison.

\begin{figure*}[t]
    \begin{subfigure}[h]{0.45\textwidth}
        \includegraphics[width=\textwidth]{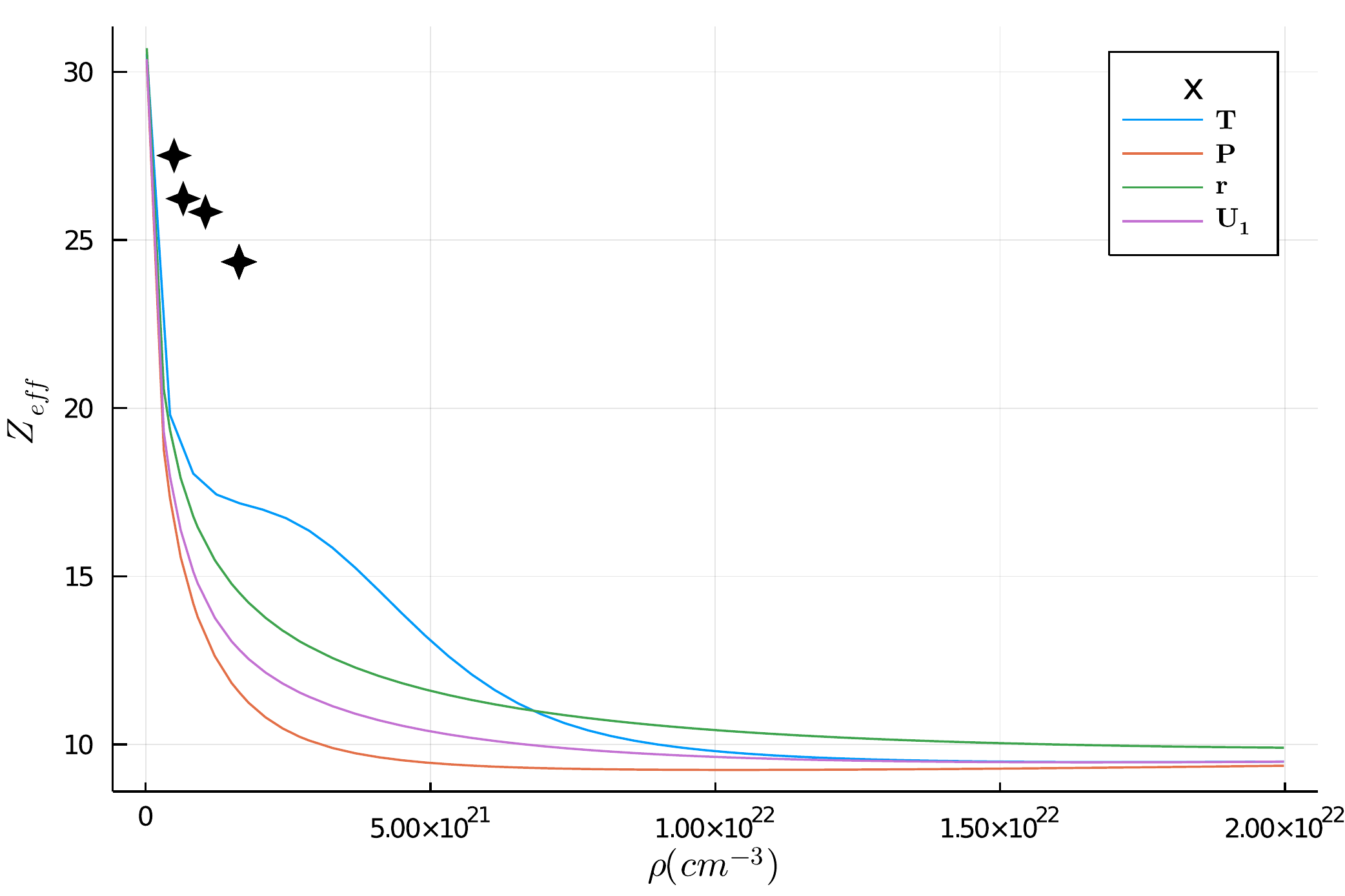}
        \caption{}
    \end{subfigure}
    \hspace{30pt}
    \begin{subfigure}[h]{0.45\textwidth}
        \includegraphics[width=\textwidth]{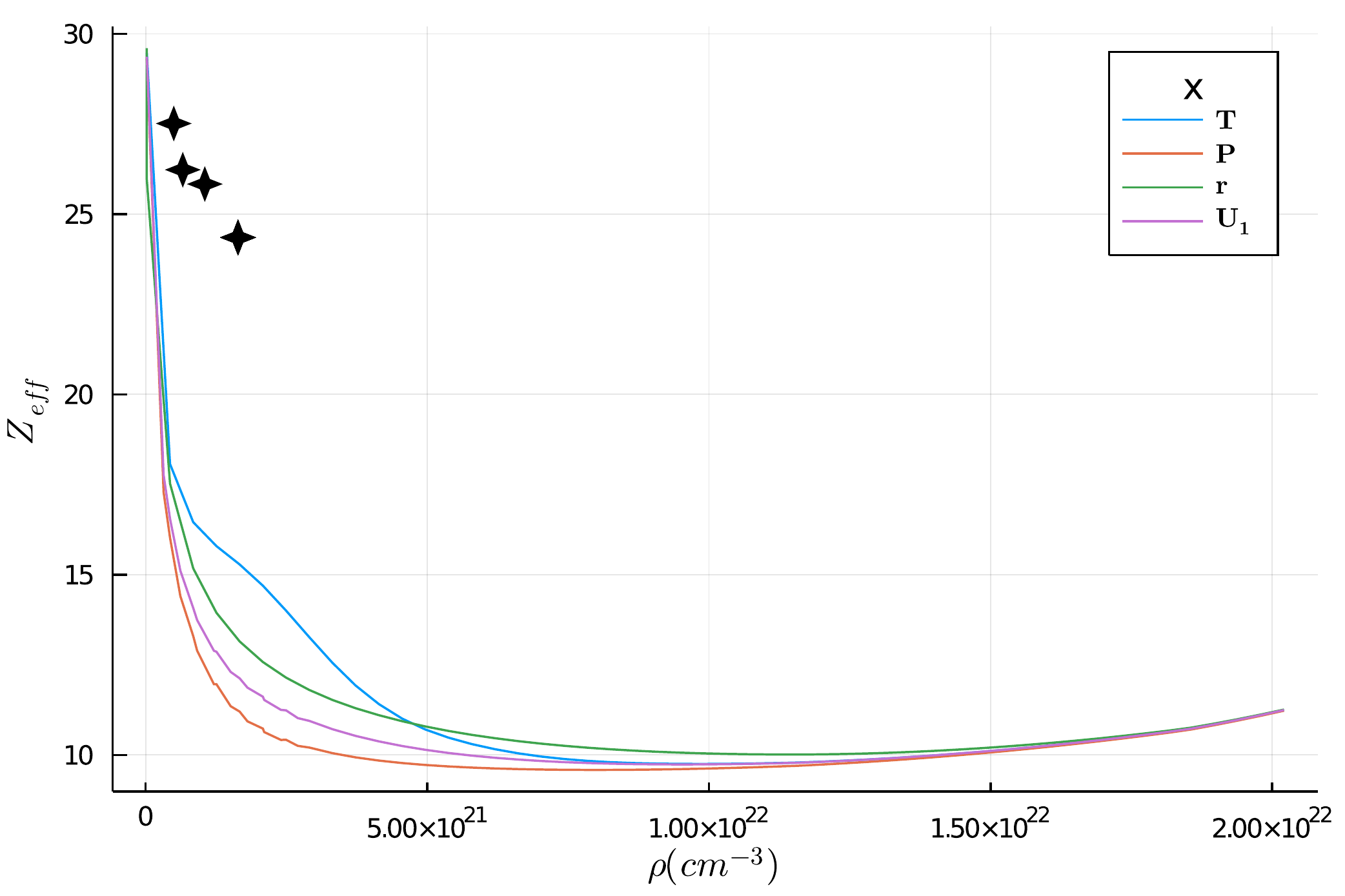}
        \caption{}
    \end{subfigure} 
    \caption{The thermal calculation of density dependence of $Z_{\text{eff}}$ for argon at 295K. (a) $Z_{\text{eff}}$ for WS radius. (b)  $Z_{\text{eff}}$ calculation with LWS radius. The different lines correspond to the different options for $U_2^x$, where x is shown in the legend. The black stars refers to the experimental data for argon at 295K\cite{tuomisaari1985localized}. \label{thermal}}
\end{figure*}
As an attempt to investigate this discrepancy between our $Z_{\text{eff}}$ and the literature data further, we calculated the averaged effective charge $\langle Z_\text{eff}\rangle$. The comparison between $\langle Z_\text{eff}\rangle$ and the literature data is shown in Figure \ref{thermal}. As suspected, because the thermal calculation using only the lowest energy orbital is not representative of the thermal average of the system, the kinetic contribution is not enough to fill the gap between our $Z_{\text{eff}}$ calculation and the literature data.

\subsection{Density dependence of $V_0$ and $Z_{\text{eff}}$ for all other noble gases \label{sectionallgas}}
 
\begin{figure*}[t] 
  \begin{subfigure}[h]{0.45\textwidth}
     \includegraphics[width=\textwidth]{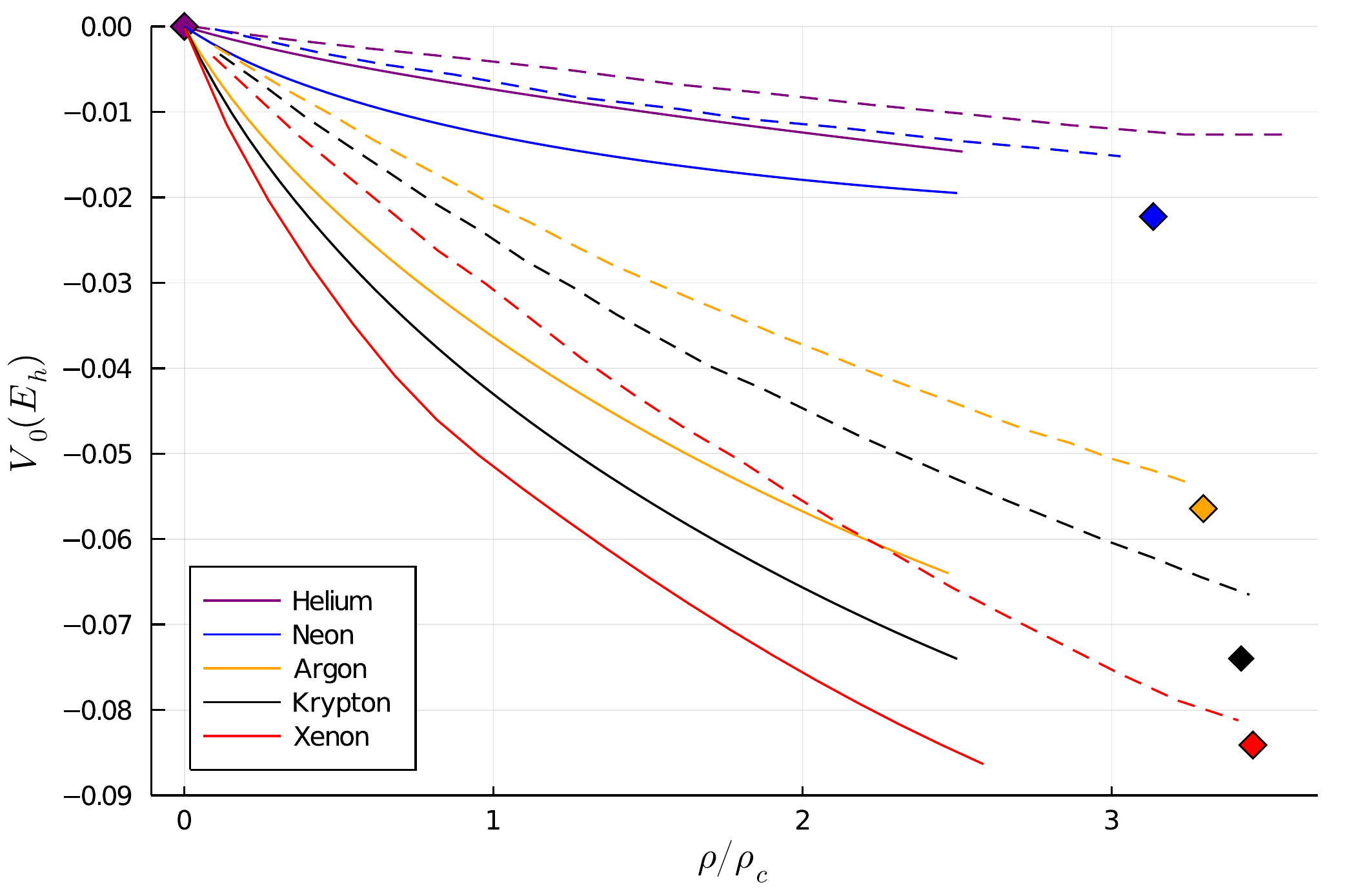}
     \caption{}
  \end{subfigure}
   \begin{subfigure}[h]{0.45\textwidth}
     \includegraphics[width=\textwidth]{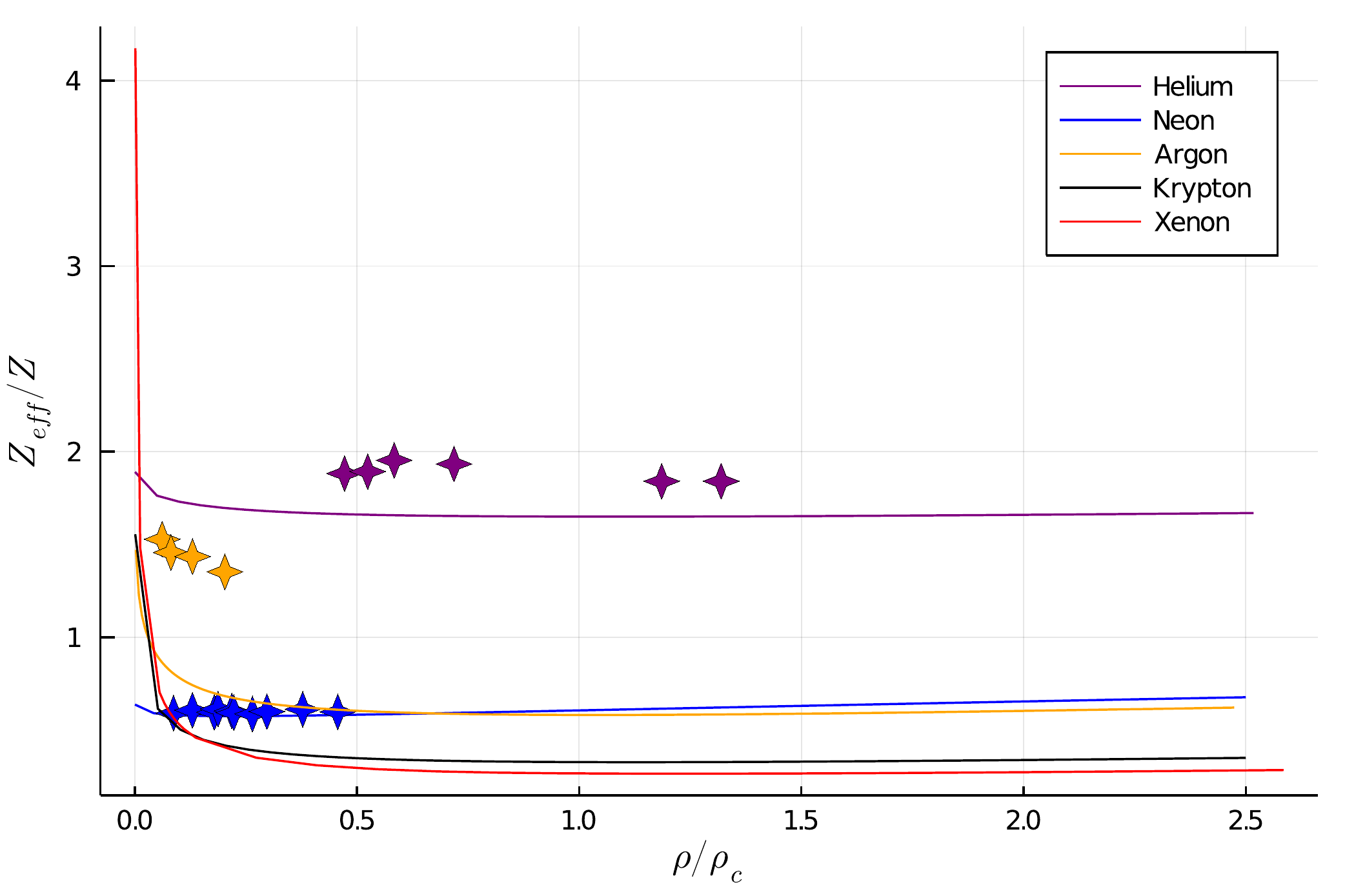}
     \caption{}
  \end{subfigure}
   \begin{subfigure}[h]{0.45\textwidth}
     \includegraphics[width=\textwidth]{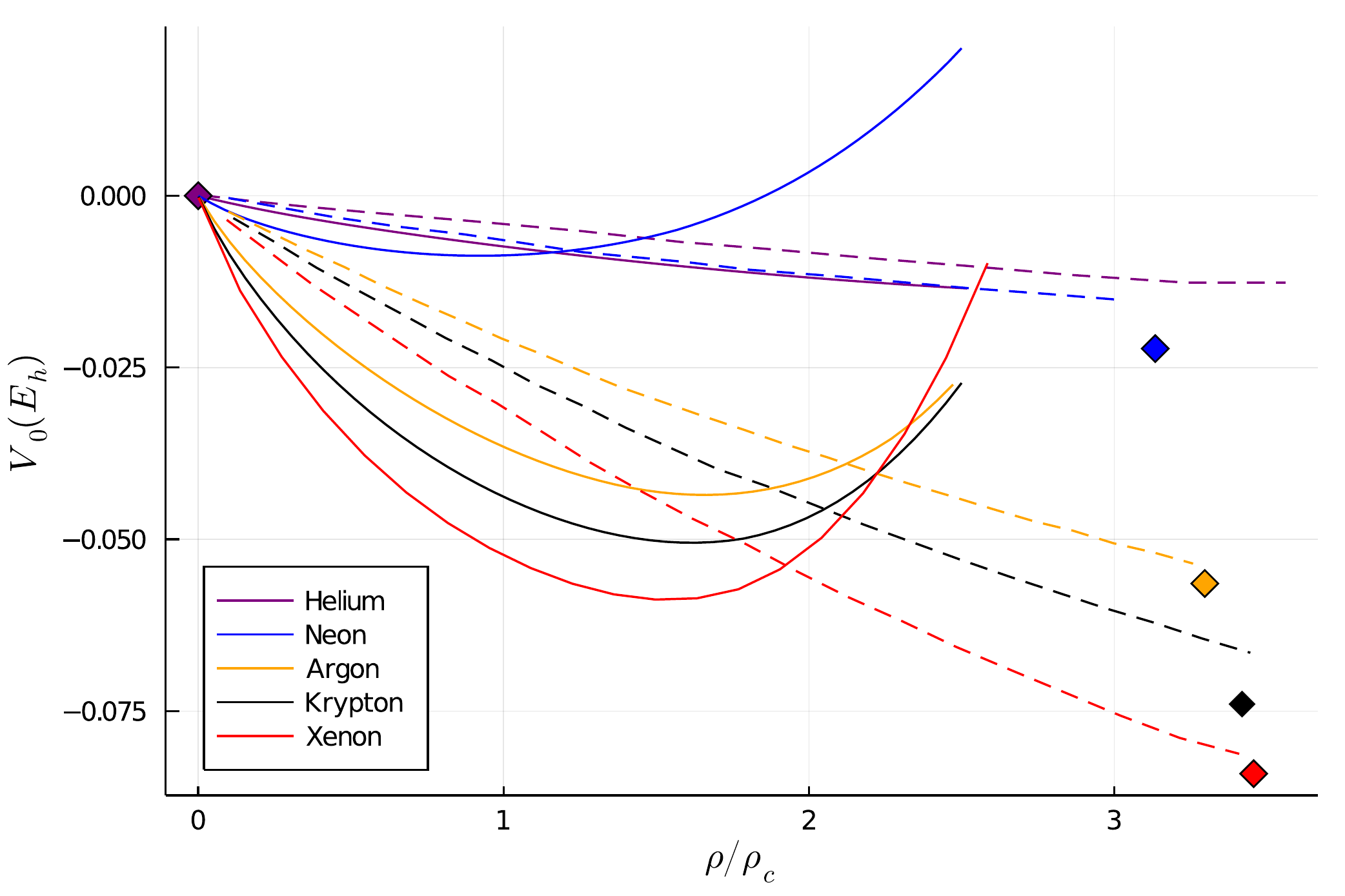}
     \caption{}
  \end{subfigure}
    \begin{subfigure}[h]{0.45\textwidth}
     \includegraphics[width=\textwidth]{UnscaledAllGasCompareZeffPaperCutoffWithU1WS.pdf}
     \caption{}
  \end{subfigure}

  \caption{The density dependence of $V_0$ and $Z_{\text{eff}}$ is shown in solid lines for all gases using WS/LWS radius: helium, neon, argon, krypton, and xenon. $\rho/\rho_c$ is the ratio between the density and the critical density, and $Z_{\text{eff}}/Z$ is the ratio between the $Z_{\text{eff}}$ and $Z$, the atomic number. (a) $V_0$ calculation using the option $U_2^{U_1}$ using WS radius (b) $Z_{\text{eff}}$ calculation using the option $U_2^{U_1}$ using WS radius. (c) $V_0$ calculation using the option $U_2^{U_1}$ using LWS radius. (d) $Z_{\text{eff}}$ calculation using the option $U_2^{U_1}$ using LWS radius. The temperatures for helium, heon, argon, krypton, xenon are 77K, 77.3K, 295K, 297K, 300K respectively. The dashed lines in (a) and (c) come from the theoretical studies using a hard-sphere potential \cite{plenkiewicz1996simple}. The diamonds in (a) and (c) refer to the experimental results obtained in crystalline samples at 4 K \cite{gullikson1988observation}. The star symbol in (b) (d) represents literature data \cite{tuomisaari1985localized,tuomisaari1988positron,canter1975positron,Nieminen1980,wright1985annihilation}. \label{WS} }
\end{figure*}
In order to demonstrate the density dependence of $V_0$ and $Z_{\text{eff}}$ for all noble gases, we present the results calculated using the option $U_2^{U_1}$ for $U_2$. We chose this method because it can be easily implemented, and it shares similarities with prior pseudo-potential approaches by introducing a clear cut-off point for the potential.

Overall, our calculations exhibit the  same qualitative trend as the hard-sphere calculation done by Plenkiewicz and co-workers \cite{plenkiewicz1996simple}. Similar with the conclusion we made with argon, their calculations are consistently higher than our calculations with $V_0$. To explore the quantitative differences, we now focus on the fluid effective scattering length, obtained using the slope of $V_0$ at low densities.

As mentioned before, the scattering length is commonly used as a metric for $V_0$ in low density regions. In table \ref{table} we show the scattering lengths calculated using PEEL and the polarized-orbital method of McEachran {\it et al} in the first and last column \cite{mceachran1979positron,mceachran1980positron,mceachran1977positron,mceachran1983elastic,mceachran2019positron}. The effective scattering lengths in the second and third column were calculated using the slope of $V_0$ as $\rho\rightarrow 0$. The comparison shows the influence of the background energy $V_0$ on the scattering lengths.

Overall, the effective scattering lengths are different from the single-atom scattering lengths, and the difference grows with the atomic weights of the elements. We also observed that for light elements, the fluid environment becomes more attractive compared to single scattering, while for heavy elements (xenon), the trend is the opposite; this is interesting. The strong deviation of the fluid effective scattering length implies that it is a poor approximation to use the single-atom scattering length in fluid calculations. We suspect that the contact potential approximation employed by equation (\ref{scatterlength}) does not hold for even dilute regions. In addition, the surrounding average assumption in the Wigner-Seitz theory is unlikely to capture elements with higher polarizability, so further deviation of the effective scattering length can occur.

In terms of the density dependence of $V_0$ and $Z_{\text{eff}}$, all noble gases show similar behaviours. The sharper drop of $Z_{\text{eff}}$ observed in argon is also observed in other noble gases. The disagreement between our calculated results and literature data is more obvious with heavier elements. Again, we suspect that the Wigner-Seitz theory averages over significant many-body interactions that are due to the large polarizability of the heavier elements. 

Similar with our observation in argon, the drastic difference in the $V_0$ calculation with WS and LWS radius is not reflected in the corresponding $Z_{\text{eff}}$ values (See Figure \ref{WS}). In order to help understand why there is lack of correspondence between the behaviours of $V_0$ and $Z_{\text{eff}}$, we also considered observing the ratios of these calculated quantities. A strong link between $Z_{\mathrm{eff}}$ and $V_{0}$ has been identified by Iakubov {\it et al} \cite{Iakubov1982}. By comparing the form of the two equations for the evaluation of these terms, Iakubov were able to show that a contact potential gives rise to the relation:
\begin{equation}
\label{eq:iakubov}
    \frac{Z_{\mathrm{eff}}}{\left(Z_{\mathrm{eff}}\right)_{\rho\rightarrow0}}\approx\frac{V_{0}/\rho}{\left(V_{0}/\rho\right)_{\rho\rightarrow0}}
\end{equation}
that is, the ratio of $Z_{\mathrm{eff}}$ to its atomic value (shown here as the limit of zero density) is equal to the ratio of the intensive quantity, background energy per density, to its atomic value.

\begin{figure*}[t] 

  \begin{subfigure}[h]{0.45\textwidth}
     \includegraphics[width=\textwidth]{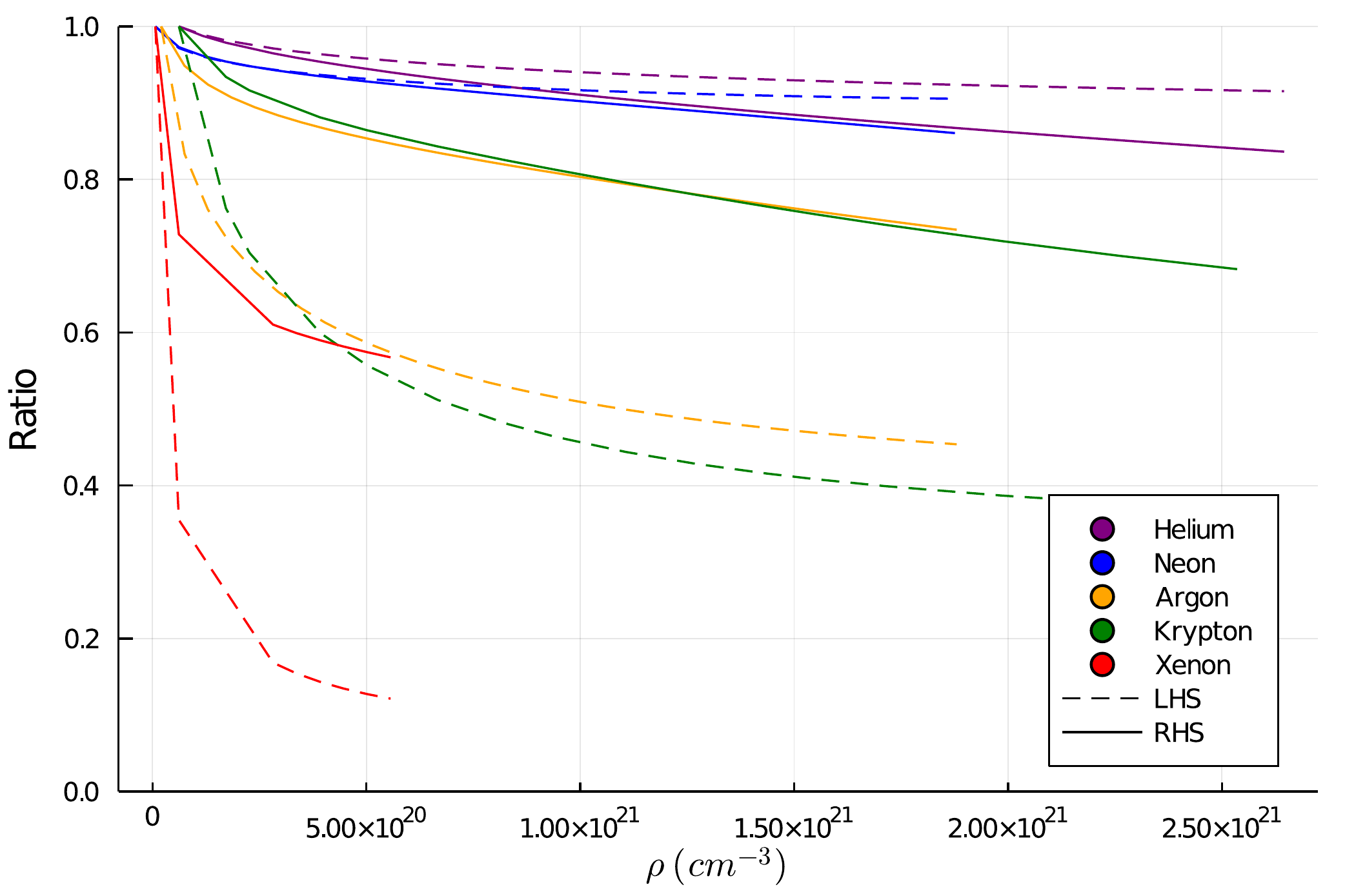}
     \caption{}
  \end{subfigure}
  \begin{subfigure}[h]{0.45\textwidth}
     \includegraphics[width=\textwidth]{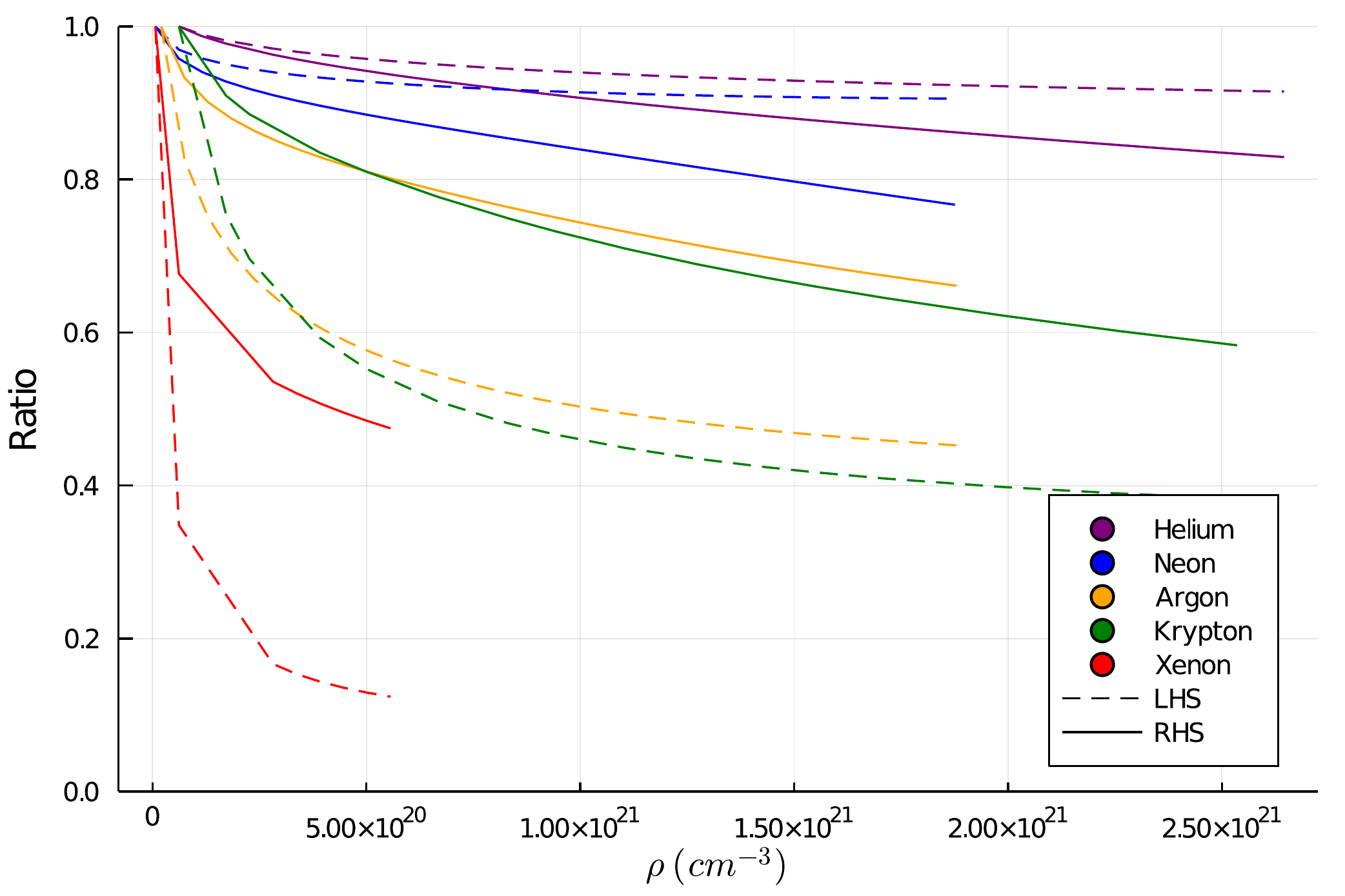}
     \caption{}
  \end{subfigure}
  \caption{\label{fig:iakubov-argon} The comparison stated by equation (\ref{eq:iakubov}) for (a) WS radius and (b) LWS radius.  The y-axis shows the ratio of the right hand side and the left hand side of the equation respectively. The different colors correspond to different noble gases, and the dashed and solid lines refer to the LHS and RHS of the equation (\ref{eq:iakubov}) respectively.}
 \end{figure*}
 
We have calculated the ratios on the left- and right-sides of equation \eqref{eq:iakubov}, and shown them in figure (\ref{fig:iakubov-argon}). For the case of WS radius, we find little agreement with equation (\eqref{eq:iakubov}) for heavy elements, while it seems to do reasonably well for light elements such as helium and neon. While in the case of LWS radius, the disagreement starts earlier (neon), but the disagreements for heavier elements seem to be lessened compared to the case with WS radius.

We believe this behaviour is the result of one or more of three factors. Either the contact potential origins of equation \eqref{eq:iakubov} cause it to be invalid for real atomic gases, or our calculations for heavy elements lack of something important, or the method in which we have calculated $Z_{\mathrm{eff}}$ is inappropriate; all three factors have been discussed above. It is our aim to resolve this discrepancy in the future by calculating an approximate transition matrix for the motion of the positron in the fluid without recourse to a surrounding average.

\begin{table}[t]
\begin{tabular}{|l|l|l|l|l|}
\hline
        & Atomic & LWS  & WS  & McEachran      \\
\hline
Helium  & -0.45             & -1.01        & -1.00        & -0.48 \\
\hline
Neon    & -0.55             & -1.55       & -1.55        & -0.61 \\
\hline
Argon   & -5.40             & -7.64       & -6.98        & -5.30 \\
\hline
Krypton & -11.21            & -11.95       & -10.39       & -10.37 \\
\hline
Xenon   & -67.80            & -14.48       & -13.42       & -45.32 \\
\hline
\end{tabular}
\caption{The first three columns were produced by PEEL. The first column is our single scattering calculation for positron-atom collision. The second and third column are our effective scattering lengths calculated in fluids for the purpose of comparing with single scattering. The last column is the single scattering calculations done by McEachran {\it et al}\cite{mceachran1979positron,mceachran1980positron,mceachran1977positron,mceachran1983elastic,mceachran2019positron}.The differences between the first and final column are due to some revised parameters in the potential calculation. \label{table}}
\end{table}

\section{Conclusion}
In this paper we present the results of different options regarding the WS calculation of $V_0$ of positron in noble gases (He, Ne, Ar, Kr, and Xe). In terms of the calculation itself, several options for the potential and WS radius were discussed. We have also commented on several options in the evaluation of $V_0$ results.

In contrast to previous theoretical calculations using  pseudo-potentials, we have employed {\it ab initio} potentials to describe the interaction between the positron and the liquid atoms, and we have shown the effects of different options for cut-offs applied for the positron-environment interaction. Among all four options proposed, the $U_2^T$ exhibits the most unusual trend. We believe that the strangeness of the cut-off $U_2^T$ is due to the over-estimation of the repulsive interaction at low densities. All other three options produce similar trends but differ in the relative positions. While $U_2^P$ consistently serves as a lower bound, the relative positions of other three options change with increasing densities. We chose $U_2^{U_1}$ as the potential option for the calculations for all noble gases, but it does not mean we believe that is the best option.

In order to evaluate our $V_0$ results, we have made several attempts. There is no direct experimental data for $V_0$ values for positron in fluids; therefore, we calculated $Z_{\text{eff}}$. This indirect comparison carries flaws as the experimentally measured $Z_{\text{eff}}$ comes from positrons with a thermal distribution, while our approach only calculates the lowest energy state of the positron. As a result, our $Z_{\text{eff}}$ results match with literature data reasonably well for light noble gases (He and Ne), while being consistently lower for heavier elements at higher densities. We implemented a preliminary calculation using a thermal distribution of positrons, but the results were not ideal. We believe the thermal calculation with only the lowest energy state is not enough, and more calculations are needed in order to properly use $Z_{\text{eff}}$ as a metric for evaluating $V_0$ calculations.

We have also tried to compare the $V_0$ calculations at low densities to the results of contact potential calculations. We have compared the effective scattering lengths calculated from slopes in $V_0$ at low densities with the scattering lengths in different gases. The difference between the two is undeniable, and it grows larger for heavier elements. The large differences between the fluid effective and single scattering lengths suggest that the contact potential approximation using the single scattering length is a poor choice for fluid calculations. The limitations of using an ensemble average for part of the potential in the Wigner-Seitz theory, we believe, further contributes to this deviation for heavier elements. We have also compared our results to the equation (\ref{eq:iakubov}), and it has shown reasonable agreement for light elements but not for heavy elements. Again, we believe the disagreement is the result of both the contact potential approximation and the surrounding average assumption of the Wigner-Seitz theory.

We have also shown the $V_0$ results using WS and LWS radius. Previously $V_0$ calculations using LWS radius for electrons were shown to be promising \cite{Evans2010}; while turning points in $V_0$ were observed using LWS radius, it is not conclusive that LWS radius is more accurate than WS radius because the corresponding $Z_{\text{eff}}$ calculations for WS and LWS radius show little difference.

In conclusion, we believe that we have used a more \emph{ab initio} potential to obtain $V_0$ for positrons in fluids, and we have examined several options for the potential calculation and the evaluation methods for $V_0$ results. For most of the potential options we provided, we have obtained reasonable agreements with indirect literature data for light elements (He and Ne), but unsatisfactory agreements for heavier elements. We believe that the Wigner-Seitz theory has limitations for calculating heavier elements with higher polarisibilities.

In future work, we will consider employing a diagrammatic expansions \cite{Polischuk1985} of the many-body interaction to describe propagating states \cite{Lax1951} directly and represent the observable quantities, such as energy, dispersion and annihilation rate. We believe this will ideally lead us to determining the most appropriate option for the ansatz in the Wigner-Seitz theory for the potential ensemble average. In terms of $Z_{\text{eff}}$ calculations, we will consider the full nonequilibrium dynamics in transport simulations to more accurately compare with experimental observations \cite{cocks2020positron}.

\section*{Data Availability}
The data that support the findings of this study are available from the corresponding author upon reasonable request.

\bibliographystyle{unsrtnat}
\bibliography{paper}
\end{document}